\documentclass[aps,prX,superscriptaddress,twocolumn]{revtex4}
\usepackage{amsmath}
\usepackage{amsfonts}
\usepackage{graphicx}
\usepackage{color}
\usepackage{dsfont}
\usepackage{verbatim}
\usepackage{mathtools}
\usepackage[normalem]{ulem}
\usepackage{comment}

\definecolor{myblue}{rgb}{.93, .93, 1}

\newcommand{\bsub}{\begin{subequations}}
	\newcommand{\esub}{\end{subequations}}

\begin{document}
	
	\title{Nonmonotonic plasmon dispersion in strongly interacting Coulomb Luttinger liquids}

	\author{Yang-Zhi~Chou}\email{yzchou@umd.edu}
\affiliation{Condensed Matter Theory Center and Joint Quantum Institute, Department of Physics, University of Maryland, College Park, Maryland 20742, USA}

\author{Sankar Das Sarma}
\affiliation{Condensed Matter Theory Center and Joint Quantum Institute, Department of Physics, University of Maryland, College Park, Maryland 20742, USA}
		
\date{\today}

\begin{abstract}
We demonstrate that the plasmon in one-dimensional Coulomb interacting electron fluids can develop a finite-momentum maxon-roton-like nonmonotonic energy-momentum dispersion. Such an unusual nonmonotonicity arises from the strongly interacting $1/r$ Coulomb potential going beyond the conventional band linearization approximation used in the standard bosonization theories of Luttinger liquids.
We provide details for the nonmonotonic plasmon dispersion using both bosonization and random-phase approximation. 
We also calculate the specific heat including the nonmonotonicity and discuss possibilities for observing the nonmonotonic plasmon dispersion in various physical systems including semiconductor quantum wires, carbon nanotubes, and the twisted bilayer graphene at sub-degree twist angles, which naturally realize one-dimensional domain-wall states.
We provide results for several different models of long-range interaction showing that the nonomonotonic charge collective mode dispersion is a generic phenomenon in one-dimensional strongly interacting electron systems.
\end{abstract}

\maketitle

\section{Introduction}

A Luttinger liquid describes the low-energy properties of one-dimensional (1D) interacting electrons which cannot be captured by the quasiparticle picture of Landau's Fermi liquid theory.
Since the interacting 1D system loses its one-to-one correspondence with the noninteracting Fermi gas, it does not manifest the discontinuity in the momentum distribution function defining a Fermi liquid \cite{Haldane1981,Voit1995,Giamarchi_Book}.
This paradigmatic non-Fermi liquid theory often assumes linear single-particle dispersion and short-range interactions for simplicity 
because bosonization techniques can then solve the interacting problem exactly. It is known that nonlinear bare single-particle energy dispersion provides higher-order irrelevant corrections to most of the critical properties of the Luttinger liquid, justifying the use of linearized models for universal properties (although detailed system specific results may very well depend on the exact bare energy dispersion). The use of a zero-range local interaction is actually less justified since the critical exponent defining the Luttinger liquid properties becomes a scale-dependent exponent in the presence of long-range interactions.

For unscreened 1D electrons interacting via the long-range 1/r Coulomb potential (e.g., semiconductor quantum wire \cite{Goni1991} and carbon nanotube \cite{Bockrath1999}), the low-energy properties are described by the Coulomb Luttinger liquid theory \cite{DasSarma1985,DasSarma1992,Schulz1993,WangCLL2001,Iucci2000} which leads to a scale-dependent velocity and a scale-dependent Luttinger parameter. In particular, a Wigner-crystal-like state with $4k_F$-correlations 
(spatially falling off slower than any power-law decay) is predicted as a universal feature of zero-temperature Coulomb Luttinger liquids \cite{Schulz1993}. This is very different from the usual short-range Luttinger liquid theory.

In a Luttinger liquid, the low-energy elementary excitations, instead of being single-particle-like as in a Fermi liquid, are in fact bosonic collective excitations with linear energy-momentum dispersion. This bosonic collective mode, often called the Tomonaga-Luttinger boson, allows the interacting fermionic problem to be studied in terms of bosonic excitations, resulting in a considerable simplification of the theory. In a Coulomb Luttinger liquid, this Tomonaga-Luttinger collective mode is nothing other than the 1D plasmon mode of the interacting electrons. (Note that the plasmon in the Coulomb Luttinger liquid contains a square root of logarithmic correction in addition to the usual linear energy-momentum dispersion \cite{DasSarma1985,DasSarma1992,Schulz1993}.) The equivalence between the 1D plasmons and the Tomonaga-Luttinger bosons is exact in a Coulomb Luttinger liquid \cite{DasSarma1992}. 

The main purpose of this work is to point out an overlooked peculiar feature in the 1D plasmon properties of the Coulomb Luttinger liquids - the existence of a nonmonotonic plasmon dispersion at finite momenta (but still much smaller than the Fermi momentum). 
For a sufficiently strong $1/r$ Coulomb interaction (or equivalently a sufficiently small Fermi velocity), we find that a maxon-roton plasmon dispersion appears at a nonuniversal finite momentum in the 1D plasmon energy. The origin of such a nonmonotonic dispersion is the long-range $1/r$ Coulomb potential.
Moreover, the spectral peak of the maxon part of the dispersion is sharp and essentially $\delta$-function-like. 
We emphasize that the low-energy long-wavelength plasmon dispersion (i.e., $\sim q\sqrt{\ln(2/q)}$, with $q$ being momentum) is not affected by the interaction strength.
Besides the plasmon dispersion, we also compute the specific heat associated with the nonmonotonic plasmon dispersion which may provide an experimental way to identify the strongly interacting regime that realizes the novel plasmon dispersion.
Establishing the nonmonotonic 1D plasmon dispersion for the strongly interacting Coulomb Luttinger liquids is the main result of this work.

Besides the 1D $1/r$ Coulomb interacting system, we find that other long-range potentials (e.g., $1/r^2$) also manifest similar finite-momentum nonmonotonic collective mode dispersion for sufficiently strong interaction. Thus, the predicted features in this work are not due to the particular form of the $1/r$ potential. Possibly, these nonmonotonic features represent the generic signatures of the strongly interacting 1D systems with long-range potentials.

The rest of the paper is organized as follows: After introducing the model in Sec.~\ref{Sec:model}, we review the basic derivations of the plasmon dispersion in Sec.~\ref{Sec:plasma_dis}. Then, we show the nonmonotonic dispersion and calculate the specific heat in Sec.~\ref{Sec:SICLL}. In Sec~\ref{Sec:Conclusion}, we discuss the possibility of observing our predicted nonmonotonicity in candidate materials (e.g., silicon quantum wire and twisted bilayer graphene at sub-degree twist angles), explain potential experimental challenges, and conclude.

\section{Model}\label{Sec:model}

We are interested in the plasmon oscillation of 1D interacting electrons. The simplest model for this is the 1D spinless fermion along the x-axis described by $\hat{H}=\hat{H}_0+\hat{H}_{I}$, where
\begin{align}
\label{Eq:H_0}\hat{H}_0=&\int dx\, c^{\dagger}(x)\left[-\frac{\partial_x^2}{2m}-\mu \right]c(x),\\
\label{Eq:H_I}\hat{H}_I=&\frac{1}{2}\int dxdx'\,  \rho(x)V(x-x')\rho(x').
\end{align}
In the above expressions, $c$ is the fermionic annihilation operator, $m$ is the electron effective mass, $\mu$ is the chemical potential, $\rho(x)=c^{\dagger}(x)c(x)$, and $V(x-x')$, the inter-particle potential, encodes the density-density interaction. 
We are particularly interested in the unscreened Coulomb interaction, characterized by $V(x)=e^2/(\kappa\sqrt{x^2+d^2})$ where $e$ is the electron charge, $\kappa$ is the dielectric constant, and $d$ is the ``transverse size'' (i.e. width of the quantum wire). 
Note that it is important to have the transverse (i.e. normal to the x-axis) dimension $d$ of the 1D system in the definition of the Coulomb interaction in order to avoid the well-known singularity of the 1D Coulomb coupling \cite{DasSarma1985}.  
For any given system the precise value of $d$ (of the order of the transverse width of the 1D system) can be calculated microscopically \cite{Lai1986}.
The Fourier transform of the potential $V(x)$ is given by $\tilde{V}(q)=\frac{2e^2}{\kappa}K_0(|q|d)$ where $K_0$ is the zero-th order modified Bessel function of the second kind. The momentum transfer due to this potential is relative to the inverse width of the quantum wire ($1/d$), which
determines the momentum scale of the nonmonotonic maxon-roton behavior of the 1D plasmon dispersion to be described below.

The spinless model given by Eqs~(\ref{Eq:H_0}) and (\ref{Eq:H_I}) describes the charge degrees of freedom in an interacting 1D system, which is of interest in the current work since plasmons are the collective charge density excitations of the system.
In the presence of spin or valley degrees of freedom, the instabilities in the spin or valley collective excitations do not affect the plasmon as long as the charge mode is decoupled from other collective degrees of freedom.
In addition, 
the degeneracy due to spin and valley will effectively enhance the density-density interaction which is typically subsumed in an overall multiplicative degeneracy factor $N$ ($N=2$ for spin degenerate systems) in the electron polarizability function. 
These complications affect only the plasmon dispersion quantitatively and will be discussed if needed. We do not further consider effects of spin or other degrees of freedom and focus on the charge density excitations (i.e. plasmons) of the 1D system (sometimes these modes are known as ``holons'').

\section{Derivation of plasmon dispersion}\label{Sec:plasma_dis}

The 1D interacting fermion Hamiltonian given by Eqs.~(\ref{Eq:H_0}) and (\ref{Eq:H_I}) can be solved by bosonization \cite{DasSarma1992,Schulz1993} after linearizing the single-particle dispersion around the Fermi energy. In particular, the bosonic mode associated with the density fluctuation is precisely the plasmon.
On the other hand, the random phase approximation (RPA) gives exactly the same energy dispersion for plasmon excitations as that obtained from bosonization \cite{DasSarma1985,DasSarma1992,DasSarma1996}. We will first review the derivation of 1D plasmon dispersion with both bosonization and RPA calculations to set a context for our predicted maxon-roton feature in the plasmon dispersion.

\subsection{Linearized theory: Coulomb Luttinger liquid}

\begin{figure}
	\includegraphics[width=0.4\textwidth]{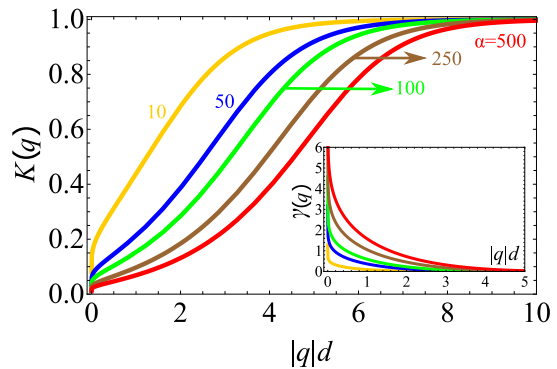}
	\caption{Scale-dependent Luttinger parameter and Luttinger exponent. We plot the Luttinger parameter $K(q)$ with different $\alpha=\frac{2e^2}{\pi v_F\kappa}$. Inset: The Luttinger exponent, $\gamma(q)=\left[K(q)+1/K(q)-2\right]/8$, is plotted. Both the $K(q)$ and $\gamma(q)$ approach to the noninteracting limit for a sufficiently large $|q|d$.
	Yellow line: $\alpha=10$; blue line: $\alpha=50$; green line: $\alpha=100$; brown line: $\alpha=250$; red line: $\alpha=500$. Note that the momentum is relative to $1/d$ rather than the Fermi wavevector $k_F$.}
	\label{Fig:LuttingerK}
\end{figure}

The interacting theory in one dimension can be studied via the standard bosonization method \cite{Shankar_Book,Giamarchi_Book}.
To incorporate the interaction, we first linearize the band in the vicinity of the Fermi energy. The physical fermion can be approximated by $c\approx e^{ik_Fx}R+e^{-ik_Fx}L$, where $R$ and $L$ denote the long-wavelength fermionic fields near the right and left Fermi points respectively in the 1D noninteracting Fermi surface.
$\hat{H}_0$ given by Eq.~(\ref{Eq:H_0}) becomes
\begin{align}\label{Eq:H_Dirac}
\hat{H}_0\rightarrow v_F\int dx\,\left[R^{\dagger}\left(-i\partial_xR\right)-L^{\dagger}\left(-i\partial_xL\right)\right],
\end{align}
where $v_F=\sqrt{2\mu/m}$ is the Fermi velocity with $\mu$ the chemical potential equal to Fermi energy at zero temperature. Since we are interested in only the long-wavelength fluctuation (i.e. momenta smaller than $k_F$, the Fermi momentum), the density operator in Eq.~(\ref{Eq:H_I}) can be expressed as $\rho\approx\rho_0= R^{\dagger}R+L^{\dagger}L$ (where we ignore the $2k_F$ and $4k_F$ components) considering only small momenta. With these approximations, the interacting fermions can be mapped to a noninteracting bosonic model. This bosonized Hamiltonian is
\begin{align}
\nonumber\hat{H}=&\int dx\,\frac{v_F}{2\pi}\left[\left(\partial_x\phi\right)^2+\left(\partial_x\theta\right)^2\right]\\
&+\frac{1}{2}\int dxdx'\,\left[\frac{1}{\pi}\partial_x\theta(x)\right]V(x-x')\left[\frac{1}{\pi}\partial_{x'}\theta(x')\right]\\
=&
\int \frac{dq}{2\pi}\frac{v(q)}{2\pi}\left[K(q)q^2\tilde{\phi}(-q)\tilde{\phi}(q)+\frac{1}{K(q)}q^2\tilde{\theta}(-q)\tilde{\theta}(q)\right],
\end{align}
where $\phi$ and $\theta$ are the phase-like and plasmonic (phonon-like) bosonic fields respectively, $v(q)$ is the scale-dependent velocity, and $K(q)$ is the scale-dependent Luttinger parameter (not to be confused with $K_0$ in the 1D Coulomb potential, which is a modified Bessel function). We can obtain the scale-dependent velocity and Luttinger parameter by solving $v(q)K(q)=v_F$ and $v(q)/K(q)=1+\frac{\tilde{V}(q)}{\pi}$. Straightforwardly,
\begin{align}
\label{Eq:vq}v(q)=&v_F\sqrt{1+\frac{\tilde{V}(q)}{v_F\pi}},\\
\label{Eq:Kq}K(q)=&\frac{1}{\sqrt{1+\frac{\tilde{V}(q)}{v_F\pi}}}.
\end{align}
Since the Luttinger parameter depends on the momentum, we might be interested in the scale-dependent Luttinger exponent as defined by $\gamma(q)=\left[K(q)+1/K(q)-2\right]/8\ge 0$.
$\gamma(q)=0$ suggests $K(q)=1$ (non-interacting limit). The scale-dependent $K(q)$ and $\gamma(q)$ are plotted in Fig.~\ref{Fig:LuttingerK}.

The equation of motion of the plasmonic boson $\theta$ can be obtained by integrating over the phase boson $\phi$ in the partition function.
The plasmonic dispersion \cite{Schulz1993} is
\begin{align}\label{Eq:omega_p_LL}
\omega_p(q)=v_F|q|\sqrt{1+\frac{\tilde{V}(q)}{v_F\pi}}=v_F|q|\sqrt{1+\frac{2e^2}{\pi v_F\kappa}K_0(|q|d)}.
\end{align}
Now, we express the above equation in dimensionless units as follows:
\begin{align}\label{Eq:dispersion}
\tilde\omega_p(\tilde{q})=\tilde{q}\sqrt{1+\alpha K_0(\tilde{q})},
\end{align}
where $\tilde\omega_p(\tilde{q})=\omega_p(q)\frac{d}{v_F}$, $\tilde{q}=|q|d$, and $\alpha=\frac{2e^2}{\pi v_F \kappa}$. The long-wavelength property is distinct from the conventional (short-range) Luttinger liquids. For $\tilde{q}\ll 1$, $\tilde{\omega}_p(\tilde{q})\approx \tilde{q}\sqrt{\ln(2/\tilde{q})}$. For $\tilde{q}\gg 1$, $\tilde{\omega}_p(\tilde{q})\approx\tilde{q}$ the same as the noninteracting dispersion.
For fermions with additional degrees of freedom (e.g., spin and valley), the dimensionless interacting parameter $\alpha$ will be enhanced by the degeneracy factor $N$. (We take $N=1$ above; for spinful fermions, $N=2$). Another important remark is that the Luttinger liquid approach can apply beyond the strict long-wavelength limit in the 1D Dirac system (e.g., edges of topological insulators and 1D domain-wall states in the twisted bilayer graphene) since the curvature is basically absent. 

Conventionally, for a quadratic single-particle dispersion, only the long-wavelength limit [i.e. Eq.~(\ref{Eq:omega_p_LL})] is considered in the literature discussing 1D plasmons and Luttinger liquid holons. 
Going beyond the long-wavelength limit where terms higher than leading order in momenta are kept in the theory, we need to worry about the finite band curvature (e.g., nonlinear Luttinger liquid theory \cite{Imambekov2012}). In the next section, we turn to a complementary approach that incorporates the band curvature exactly.

\subsection{Random phase approximation}

In addition to the bosonization approach, we study the problem using the RPA. In fact, the 1D plasmon dispersion was originally computed using the RPA a long time ago \cite{DasSarma1985}, and much theoretical work has been done on 1D plasmon dispersion using the RPA in the context of studying collective modes in semiconductor quantum wires \cite{DasSarma1996,Li1989,Li1991}.
In contrast to the band linearization approximation in bosonization, we keep the full quadratic single-particle dispersion and treat the interaction effect via the summation of the infinite series of the ``bubble'' diagrams.
We emphasize that the 1D RPA plasmon mode reproduces the exact bosonization long-wavelength result, but the RPA enables us to go beyond the long-wavelength linearized limit. 

The standard dynamic dielectric function in RPA is given by
\begin{align}\label{Eq:dielectric}
\epsilon(\omega,q)=1-\tilde{V}(q)\Pi_0(\omega,q),
\end{align}
where the irreducible polarization function \cite{DasSarma1985}
\begin{align}
\Pi_0(\omega,q)=\frac{m}{2\pi|q|}\ln\left[\frac{\omega^2-\omega^2_-}{\omega^2-\omega^2_+}\right],
\end{align}
and $\omega_{\pm}=v_F|q|\pm\frac{q^2}{2m}=v_F|q|\pm\frac{v_F}{2k_F}q^2$. The plasmon corresponds to the zero of the dielectric function $\epsilon(\omega,q)$ [given by Eq.~(\ref{Eq:dielectric})]. The plasmon dispersion in RPA is then
\begin{align}\label{Eq:dispersion_RPA}
\omega_p(q)=\left[\frac{A(q)\omega_+^2-\omega_-^2}{A(q)-1}\right]^{\frac{1}{2}},
\end{align}
where $A(q)=\exp\left[\frac{2|q|\pi}{k_F\tilde{V}(q)/v_F}\right]$. Equation (\ref{Eq:dispersion_RPA}) is consistent with the results in Ref.~\cite{DasSarma1985} when setting the spin degeneracy factor $N=1$ (spinless case). In the limit $|q|/k_F\ll 1$, we expand $\omega_p(q)$ to $O(q^2/k_F^2)$ and recover the Luttinger liquid result \cite{Schulz1993} given by Eq.~(\ref{Eq:omega_p_LL}). The plasmon dispersion in the large $|q|d$ limit [$A(q)\gg 1$] is precisely the upper bound of the particle-hole continuum. 
Note that the long-wavelength limit in the formal many-body theory corresponds to $q\ll k_F$ whereas the 1D plasmon dispersion depends also on an independent dimensionless parameter $qd$ in addition to the parameter $q/k_F$. Thus, the 1D plasmon dispersion depends on two independent length parameters: $d$ and $1/n$, where $n$ is the 1D electron density in the system defining $k_F=\pi n/N$.

\section{Strongly interacting Coulomb Luttinger liquid}\label{Sec:SICLL}

\begin{figure}
	\includegraphics[width=0.4\textwidth]{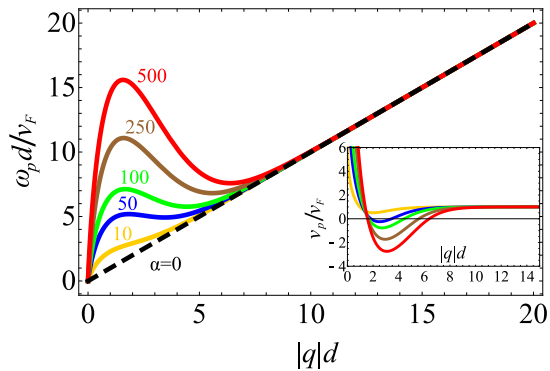}
	\caption{Plasmon dispersion in the Coulomb Luttinger liquids. The plasmon dispersion given by Eq.~(\ref{Eq:dispersion}) with different $\alpha=\frac{2e^2}{\pi v_F \kappa}$. Inset: The plasmon velocity, $v_p\equiv\frac{d\omega_p(q)}{dq}$, is plotted. The zeros of $v_p$ identify the maxon and roton wavevectors.
		Yellow line: $\alpha=10$; blue line: $\alpha=50$; green line: $\alpha=100$; brown line: $\alpha=250$; red line: $\alpha=500$. The black dashed line ($\alpha=0$) corresponds to a noninteracting 1D linearly dispersing fermion and sets the upper bound of the particle-hole continuum. The plasmon dispersion develops a roton like minimum when $\alpha>33$ (estimated numerically). Note that the momentum is relative to $1/d$ rather than the Fermi wavevector $k_F$.}
	\label{Fig:dispersion}
\end{figure}

\begin{figure}
	\includegraphics[width=0.4\textwidth]{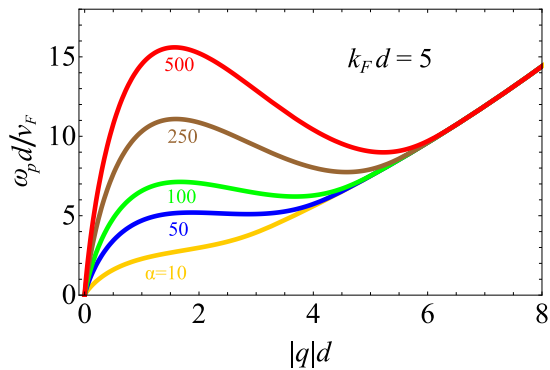}
	\caption{Plasmon dispersion of 1D Coulomb interacting fermions with RPA. We plot Eq.~(\ref{Eq:dispersion_RPA}) with different $\alpha=\frac{2e^2}{\pi v_F \kappa}$. 
		Yellow line: $\alpha=10$; blue line: $\alpha=50$; green line: $\alpha=100$; brown line: $\alpha=250$; red line: $\alpha=500$. For $|q|d>6$, the plasmon dispersion merges with the upper bound of the particle-hole continuum.
		We have set $k_Fd=5$ for all of the curves. Note that the momentum is relative to $1/d$ rather than the Fermi wavevector $k_F$.}
	\label{Fig:dispersion_RPA}
\end{figure}

\begin{figure}
	\includegraphics[width=0.4\textwidth]{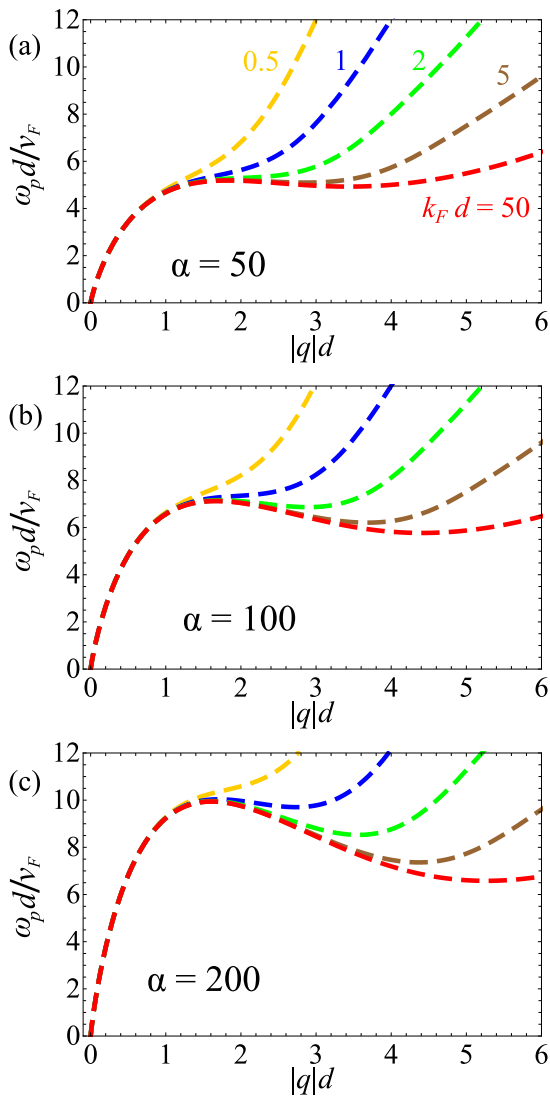}
	\caption{Plasmon dispersion using RPA with varying $k_Fd$. 
		We plot Eq.~(\ref{Eq:dispersion_RPA}) with various $\alpha$ [(a) $50$, (b) $100$, and (c) $200$] and different values of $k_Fd$. 
		Yellow dashed line: $k_Fd=0.5$; blue dashed line: $k_Fd=1$; green dashed line: $k_Fd=2$; brown dashed line: $k_Fd=5$; red dashed line: $k_Fd=50$. Note that the momentum is relative to $1/d$ rather than the Fermi wavevector $k_F$.}
	\label{Fig:dispersion_RPA_KF}
\end{figure}

Most of the low-energy properties of the Coulomb Luttinger liquids have been studied systematically \cite{DasSarma1992,Schulz1993,WangCLL2001}. In particular, the ground state develops a Wigner-crystal-like quasi long-range order with slowly spatially decaying (slower than any power-law) $4k_F$ oscillations regardless of the interaction strength \cite{Schulz1993}. Therefore, the Coulomb Luttinger liquids cannot in general be characterized by a single exponent as in the short-range case, and the Luttinger parameter is now scale-dependent \cite{WangCLL2001}. Here, we show that ``strongly'' interacting Coulomb Luttinger liquids can develop a nonmonotonic plasmon dispersion. Concomitantly, the specific heat in the strongly interacting regime develops a low-temperature suppression arising directly from the plasmon nonmonotonicity.

\subsection{Non-monotonic plasmonic dispersion}

\begin{figure}
	\includegraphics[width=0.4\textwidth]{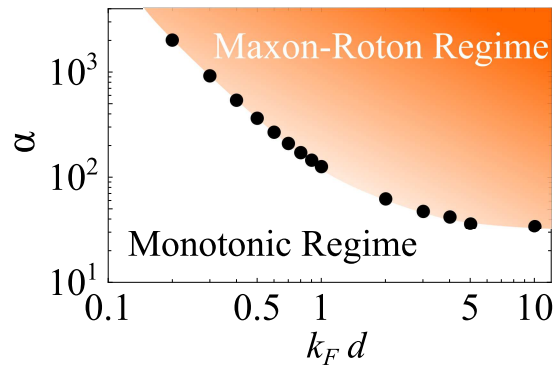}
	\caption{Diagram of qualitatively distinct plasmon dispersion regimes using RPA calculations. The black dots represent
		the interaction threshold ($\alpha_c$) for realizing nonmonotonic plasmon dispersion given by Eq.~(\ref{Eq:dispersion_RPA}). For $\alpha>\alpha_c$ (orange shaded regime), the maxon-roton feature is manifest. The value of $\alpha_c$ decreases when $k_Fd$ increases.
		The value of $\alpha_c$ using RPA saturates at $\sim 33$ (the $\alpha_c$ obtained from the Luttinger liquid dispersion) for $k_Fd\gg 1$.}
	\label{Fig:alphaC}
\end{figure}

The plasmon dispersion derived in the previous section has been verified in a number of experiments \cite{Goni1991,Demel1991,Strenz1994}. We can simply take Eq.~(\ref{Eq:dispersion}) and insert the material-dependent interaction parameter $\alpha$, which depends only on the background dielectric constant ($\kappa$) and the bare Fermi velocity.  In principle, therefore, $\alpha$ can be tuned by changing the electron density (e.g., through gating) and/or by changing the substrate to modify the dielectric background.  Since $v_F=\pi n/(N m)$, one can enhance $\alpha$ simply by decreasing the 1D carrier density through external gating.  

A natural question arises: can a larger $\alpha$, corresponding to a stronger interacting system, give a qualitatively different plasmon dispersion?
To answer the above question, we examine Eq.~(\ref{Eq:dispersion}) by varying $\alpha=\frac{2e^2}{\pi v_F \kappa}$. 
In Fig.~\ref{Fig:dispersion}, the calculated plasmon dispersion develops an enhanced peak and a local minimum for a sufficiently large $\alpha>\alpha_c$, where $\alpha_c$ is the critical value of the dimensionless interaction parameter for producing this nonmonotonicity. The nonmonotonic dispersion is reminiscent of the famous roton-maxon dispersion in liquid helium.
Large $\alpha$ corresponds to either a strong interaction (i.e. small $\kappa$) or a small Fermi velocity (low carrier density). To identify $\alpha_c$, we compute the derivative of the plasmon dispersion with varying $\alpha$. The condition of getting zero plasmon velocity [$v_p(q)=d\omega_p(q)/dq=0$] corresponds to $\alpha\left[\frac{\tilde{q}}{2}K_1(\tilde{q})-K_0(\tilde{q})\right]=1$. 
We find that $\alpha_c\approx 33$ numerically using the linearized Luttinger liquid theory [i.e. Eq.~(\ref{Eq:dispersion})]. In addition, the maxon and roton wavevectors are around $2/d$ (i.e. where the plasmon dispersion nonmonotonicity occurs). Note that the nonmonotonicity disappears for $\alpha<\alpha_c$, explaining why earlier work missed the maxon-roton structure in the plasmon dispersion since the regime of large $\alpha$ (small $\kappa$ and small $v_F$) was never studied before in the context of 1D plasmon dispersion theories.

An important question is the stability of this predicted dispersion in finite-wavevector Coulomb Luttinger liquid calculations, especially the robustness against the finite curvature. To answer this, we check Eq.~(\ref{Eq:dispersion_RPA}) which follows from the full quadratic band using the full RPA theory. 
In Figs.~\ref{Fig:dispersion_RPA} and \ref{Fig:dispersion_RPA_KF}, we show that the nonmonotonic plasmon dispersion persists but with a	 value of $\alpha_c$ strongly depending on $k_Fd$. 
In Fig.~\ref{Fig:alphaC}, we show that $\alpha_c$ monotonically decreases as a function of $k_Fd$ and saturates at $\sim 33$ [the same $\alpha_c$ as obtained from the Luttinger liquid dispersion in Eq. (\ref{Eq:dispersion})] for $k_Fd\gg 1$.
We conclude that the nonmonotonic 1D plasmon dispersion does survive in the presence of the band curvature effects, and both the standard bosonized Luttinger liquid theory and the full RPA give the maxon-roton plasmon dispersion at finite momenta ($q \sim  2/d$) provided the Coulomb coupling exceeds a critical value.
As presented in Figs.~\ref{Fig:dispersion_RPA_KF} and \ref{Fig:alphaC}, a larger value of $k_Fd$ indeed enhances the maxon-roton feature in the full RPA plasmon dispersion. Thus, the maxon-roton feature, arising in the strongly interacting system, is not an artifact of the linearized Luttinger liquid theory as it exists in the full RPA theory too except that in the full RPA, the nonmonotonicity depends both on the dimensionless Coulomb coupling strength ($\alpha$) and the dimensionless Fermi momentum ($k_Fd$). 

The maxon (peak) and roton (local minimum) wavevectors depend on the width of the quantum wire ($d$) instead of the Fermi wavevector. With a wide quantum wire (i.e. large $d$), the maxon and roton wavevectors can be much smaller than the Fermi wavevector. This suggests that the nonmonotonicity in the plasmon dispersion comes from the ``low-energy'' properties of the Coulomb Luttinger liquids since it can occur already for $q\ll k_F$. Mathematically, the long-wavelength 1D plasmon dispersion $\omega_p(q)\approx v_Fq\sqrt{\ln[2/(qd)]}$ already gives the nonmonotonic dispersion, clearly showing that the maxon-roton feature is inherent in the long-range nature of the Coulomb coupling which is inherently dependent on the cutoff scale ``$d$.''
Moreover, the nonmonotonic part of the plasmon dispersion (especially the maxon peak) is always well above the 1D particle-hole continuum and is undamped by particle-hole pair creations. 
We therefore expect to observe a sharp essentially delta-function-like spectral peak in the dynamical structure factor corresponding to the maxon-roton dispersion feature which should clearly show up in inelastic scattering experiments such as light scattering and phtoemission spectroscopies.

An important remark is that the scale-dependent Luttinger parameter $K(q)$ is always monotonically increasing to 1 with increasing $q$ (see Fig.~\ref{Fig:LuttingerK} with no nonmonotonicity whatsoever). This implies that the nonmonotonic plasmon dispersion does not change the well-known zero-temperature Wigner-crystal-like phase of the 1D Coulomb systems \cite{Schulz1993,Vu2018}. Nevertheless, we might expect that the nonmonotonicity of the plasmon dispersion contributes to the finite-temperature properties where higher energy excitations may contribute.
In the next section, we turn to the specific heat which provides a way to identify the strongly interacting regime (the same regime of $\alpha$ that realizes the nonmonotonic plasmon dispersion).

\subsection{Specific heat}

When the plasmon oscillation develops a nonmonotonic dispersion, the density of states of the plasmon diverges at the maxon and roton energies. In addition, the dispersion $\omega_p(q\rightarrow 0)\sim q\sqrt{\ln[2/(qd)]}$ implies a vanishing low-energy density of states.
These singular features will be reflected in the thermodynamic quantities like specific heat which we discuss in this section.

\begin{figure}
	\includegraphics[width=0.4\textwidth]{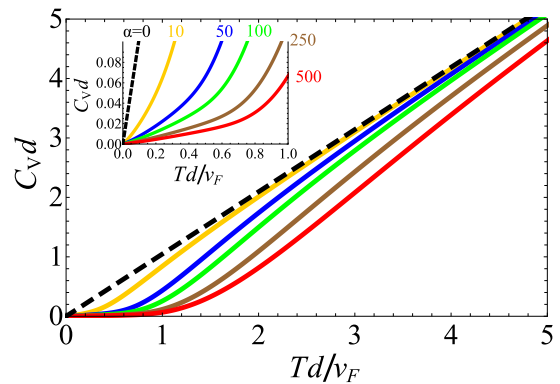}
	\caption{Specific heat of the Coulomb Luttinger liquids with various values of $\alpha$. The low-temperature suppression in the specific heat depends on the value of $\alpha$ and therefore can identify the regime realizing nonmonotonic plasmon dispersion. Black dashed line: $\alpha=0$; yellow solid line: $\alpha=10$; blue solid line: $\alpha=50$; green solid line: $\alpha=100$; brown solid line: $\alpha=250$; red solid line: $\alpha=500$.}
	\label{Fig:Cv}
\end{figure}

The specific heat can be computed straightforwardly based on the plasmon dispersion in Eq.~(\ref{Eq:dispersion}). The expression for the specific heat is
\begin{align}\label{Eq:Cv}
C_V=&\left(\frac{v_F}{Td}\right)^2\frac{1}{\pi d}\int_{0}^{\infty}d\tilde{q}\,\frac{\left[\tilde{\omega}_p(\tilde{q})\right]^2e^{\frac{v_F}{T d}\tilde{\omega}_p(\tilde{q})}
}{\left[e^{\frac{v_F}{T d}\tilde{\omega}_p(\tilde{q})}-1\right]^2},
\end{align}
where $\tilde\omega_p(q)$ is the dimensionless plasmon dispersion in Eq.~(\ref{Eq:dispersion}). The specific heat results given by Eq.~(\ref{Eq:Cv}) are plotted in Fig.~\ref{Fig:Cv}. In the high-temperature limit, the specific heat $C_V\approx \frac{\pi T}{3v_F}$, which is consistent with what is expected for the noninteracting 1D spinless fermions \cite{Giamarchi_Book}. In the inset of Fig.~\ref{Fig:Cv}, the low-temperature specific heat shows a suppression which depends strongly on $\alpha$. Although all of the low-temperature specific heat curves roughly follow a linear temperature dependence, the prefactor gets smaller for larger $\alpha$. 
We find that the low-temperature specific heat is significantly reduced for $\alpha \gg \alpha_c$. Such a reduction of the low-temperature specific heat is caused by the small density of states of the low-energy plasmon for large $\alpha$ where the maxon-roton dispersion feature arises. It is important to note that the low-temperature specific heat cannot be described by an exponential Arrhenius scaling but appears to be close to linear in temperature.  
This is consistent with the gapless nature of the plasmon in one dimension. However, the precise low-temperature form is not known analytically except that we find that it is significantly suppressed for strong interaction. This strong suppression for large $\alpha$ is directly connected with the nonmonotonic maxon-roton plasmon dispersion.

The results in this section are only on the charge sector contribution to the specific heat. There may be contributions to the specific heat from spinons and other excitations outside the charge sector.
Nevertheless, we expect a large suppression of the low-temperature specific heat for strongly interacting Coulomb Luttinger liquids as shown in Fig.~\ref{Fig:Cv} since the contribution from plasmons will indeed be significantly suppressed at strong coupling.

\section{Conclusion}\label{Sec:Conclusion}

We have theoretically demonstrated that a Coulomb Luttinger liquid can manifest a finite-momentum nonmonotonic 1D plasmon dispersion in the strongly interacting regime, preserving still the low-energy charge collective mode behavior [i.e. $\omega_p(q\rightarrow 0)\sim q\sqrt{\ln[2/(qd)]}$] at long wavelengths. 
We emphasize that our results are model-independent: both Luttinger liquid theory (for linearized energy dispersion) and the RPA (for parabolic energy dispersion) give the nonmonotonic plasmon dispersion at finite wavenumbers.
We also show that the specific heat of the Coulomb Luttinger liquids depends strongly on the interaction strength ($\alpha$). In particular, the magnitude of the low-temperature specific heat suppression can identify the strongly interacting regime, the same regime realizing the nonmonotonic plasmon dispersion. 

Where to look for the nonmonotonicity in the 1D plasmon dispersion is an important question, which we can answer only partially. One needs a large $\alpha\sim 1/(v_F\kappa)$ (since $\alpha>\alpha_c$ is necessary) and a large $d$ (since $q\sim1/d$ is necessary). These two conditions are in principle independent since the experimental 1D system has four independent physical parameters: $n$ (carrier density) and $m$ (effective mass) determining the Fermi velocity, $\kappa$ (the background dielectric constant) determining the strength of Coulomb coupling, and $d$ (the transverse width) determining the cut off for the Coulomb divergence. 
We emphasize that, as long as $q\ll k_F$ (which is allowed since $k_F$ and $d$ are independent physical parameters), our theory (both RPA and Luttinger liquid theory) are essentially exact, and hence, our predictions are independent of our approximation schemes.
The ideal 1D system should have very low $n$ (and not too small an $m$) so that $v_F$ is small along with a rather small $\kappa$ so that the Coulomb coupling is strong. In addition, $d$ should not be too small or large, so that the characteristic wave number, $q \sim 1/d$, where the nonmonotonicity occurs is neither too large nor too small.  We estimate 1D quantum wires made of silicon with a low carrier density to be the possible system for the observation of our predicted maxon-roton feature, but other 1D systems should be tunable to the interesting regime since the number of independent tunable parameters ($n$, $m$, $d$, $\kappa$) is twice the number of conditions ($\alpha>\alpha_c$ and $q>1/d$) necessary for the predicted physics to occur.

One problem to worry about is the constraint $k_Fd >1$ which also seems to be crucial for the manifestation of the maxon-roton nonmonotonicity in $k^2$ dispersing fermion bands [i.e. Eq.~(\ref{Eq:H_0})]. 
In Fig. \ref{Fig:dispersion_RPA_KF}, we show the RPA calculated 1D plasmon dispersion for $\alpha=50$, $100$, $200$ and $k_Fd=0.5$, $1$, $2$, $5$, $50$.  As is clear from these plots, the nonmonotonicity is favored when $k_Fd>1$ as well as $\alpha \gg1$.
The interaction threshold $\alpha_c$ (the minimum value of $\alpha$ realizing nonmonotonicity) versus $k_Fd$ is plotted in Fig.~\ref{Fig:alphaC}. We show that $\alpha_c$ decreases as a function of $k_Fd$ and saturates at the linearized Luttinger liquid value (i.e. $\alpha_c\approx33$) for $k_Fd\gg 1$.
Unfortunately, $k_Fd>1$ and $\alpha \gg1$ are somewhat mutually exclusive in doped parabolic 1D electron systems such as semiconductor quantum wires since $v_F=k_F/m$, and $\alpha \sim 1/v_F$.  Thus, a small $v_F$ (to make $\alpha$ large) typically implies a small $k_F$ as well which makes it difficult to satisfy the condition $k_Fd>1$. In addition, $k_Fd>1$ condition is in conflict with the system being in the strictly 1D limit as the higher quantized levels in the system may become relevant. Even in such a situation, however, the main charge collective mode will manifest a charge density oscillation reflecting the total charge density, and the plasmon would remain 1D character as long as the inter-level transitions are suppressed. A complete theory including the multilevel situation is well outside the scope of the current work and should follow the formalism of Refs.~\cite{Lai1986Plasmon,Li1991,Li1990}.  Our current work only points out that in the strongly interacting (i.e. $\alpha\gg 1$) situation, the charge collective bosonic mode (i.e. 1D plasmon) in a Coulomb Luttinger liquid develops a nonmonotonic energy-momentum dispersion.  The actual experimental manifestation of the physics discussed in this work may be more likely in a lattice system (rather than in a doped system) where the Fermi velocity and the Fermi momentum may not be directly connected with the Fermi velocity arising from band dispersion (and can be very small in flat bands) and the Fermi momentum arises from the band filling, thus enabling both $\alpha\gg 1$ and $k_Fd>1$ conditions to be simultaneously satisfied. The complications stated above explain why the plasmon nonmonotonicity has not been observed in literature.
The ideal system manifesting the nonmonotonicity is a 1D system where the bare band dispersion is linear to very large momenta (i.e. very small band curvature) so that the complication arising from $k_Fd>1$ is simply not relevant. In this sense, graphene-related systems (carbon nanotubes, graphene nanoribbons, twisted bilayer graphene, etc.) may turn out to be the ideal place to look for the maxon-roton physics predicted in our work.

\begin{figure}
	\includegraphics[width=0.375\textwidth]{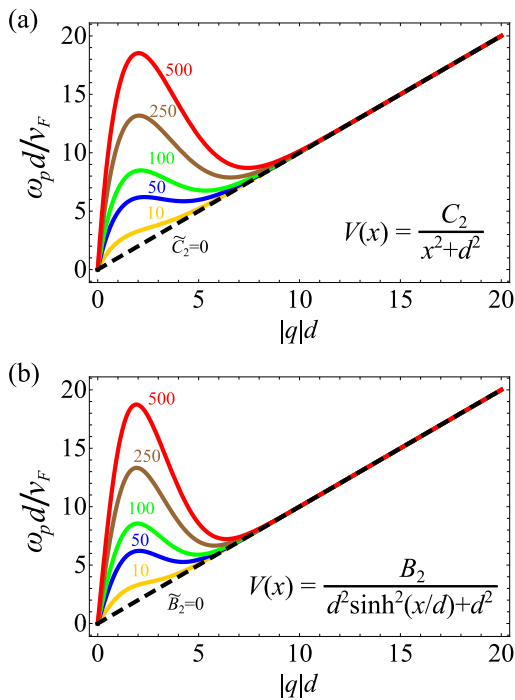}
	\caption{
		Plasmon dispersion of 1D fermions with different long-range potentials. (a) Inverse square potential, $V(x)=C_2/(x^2+d^2)$. The dispersion is given by $\omega_p d/v_F=\tilde{q}\sqrt{1+\tilde{C}_2\tilde{q}^{1/2}K_{1/2}(\tilde{q})}$, where the dimensionless strength $\tilde{C}_2=\frac{C_2\sqrt{2\pi}}{\pi v_F d}$, $K_{1/2}$ is the modified Bessel function of the second kind, and $\tilde{q}=|q|d$. The plasmon dispersion develops a maxon-roton feature when $\tilde{C}_2>33$ (estimated numerically). (b) $V(x)=B_2/\left[d^2\sinh(x/d)+d^2\right]$.
		The dispersion is given by $\omega_p d/v_F=\tilde{q}\sqrt{1+\tilde{B}_2\tilde{q}^{1/2}\text{csch}(\tilde{q}\pi/2)}$, where the dimensionless strength $\tilde{B}_2=\frac{B_2 \pi}{\pi v_F d}$. The plasmon dispersion develops a maxon-roton feature when $\tilde{B}_2>21$ (estimated numerically).
		Black dashed line: $C_2=0$ ($B_2=0$); yellow line: $C_2=10$ ($B_2=10$); blue line: $C_2=50$ ($B_2=50$); green line: $C_2=100$ ($B_2=100$); brown line: $C_2=250$ ($B_2=250$); red line: $C_2=500$ ($B_2=500$). All the curves are calculated via Luttinger liquid approximation.
	}
	\label{Fig:other_dispersion}
\end{figure}

Interestingly, the recent experiments in the twisted bilayer graphene \cite{tbg1,tbg2,Yoo2019_reconstruction,Huang2018Helical_TBLGexp,Xu2019_GiantOscillation} may provide a new platform for investigating the plasmon nonmonotonicity. When the twist angle is smaller than 1 degree, the 1D domain-wall structure (which separates the effectively AB and BA stacking regions) is manifest \cite{Yoo2019_reconstruction,Huang2018Helical_TBLGexp,Xu2019_GiantOscillation,Efimkin2018}. It is important to emphasize that the 1D domain-wall states, which can be well described by the 1D Dirac fermion with linearized dispersion [i.e. Eq.~(\ref{Eq:H_Dirac})], are qualitatively different from the conventional $k^2$ dispersing band. Thus, the plasmon dispersion from bosonization [i.e. Eq.~(\ref{Eq:omega_p_LL}) which is ignorant of $k_F$] captures the plasmon in such 1D domain-wall states. The domain-wall states contain four counter-propagating helical pairs, arising as the boundary separating quantum valley Hall state in the AB (winding number $\nu=1$) and BA ($\nu=-1$) staking registries \cite{Zhang2013,Efimkin2018}. The width of the domain-wall states (i.e. $d$) is estimated to $2-3$nm (based on lattice relaxed calculations in \cite{Guinea2019}).
The nonmonotonic plasmon dispersion may be observed in the twisted bilayer graphene close to the second or the third magic angles \cite{Bistritzer,Tarnopolsky2019} because the domain-wall state velocity in the miniband is significantly quenched (i.e. large enhancement in $\alpha$). Although the characteristic wave number for nonmonotonicity is considerably large ($\sim1/d$), the long-wavelength part (i.e. smaller than $1/d$) of plasmon dispersion is boosted up by a factor $\sqrt{\alpha}$ (with $\alpha\gg 1$), enabling a large energy separation between plasmon dispersion and the particle-hole continuum. 
Such a system with a large energy mismatch, resulting in an undamped plasmon dispersion, is very similar to the recent study on the surface plasmon in twisted bilayer graphene at a magic angle \cite{Lewandowski201909069} although it approaches the problem from the two-dimensional limit rather than from the 1D limit (this work). We further predict that the specific heat has a strong low-temperature suppression in the same regime manifesting the plasmon dispersion nonmonotonicity. The precise values of the magic angles lower than 1 degree (the first magic angle) are still openly debated \cite{Bistritzer,Tarnopolsky2019,Carr2019}. Nevertheless, the nonmonotonic plasmon dispersion predicted in this work does not rely on the magic angle fundamentally as long as the domain-wall velocity is small enough and the 1D approximation applies.

Theoretically, we can ask if a different interaction potential (but still long-ranged) also realizes the nonmonotonic dispersion. The answer is in the affirmative. The same analysis as in the current work for other long-range potentials manifests similar nonmonotonicity in the collective mode dispersion in the strong coupling regime, showing our finding of the maxon-roton feature to be \textit{universal} for 1D long-range interacting systems.
For example, we check a power-law potential $V(x)=C_n/(\sqrt{x^2+d^2})^n$ for $n>0$ and find nonmonotonic collective mode dispersion for a sufficiently large $C_n$. {We plot the plasmon dispersion for $n=2$ in Fig.~\ref{Fig:other_dispersion}(a). For $n=2$, the model can be viewed as the continuum version of the Haldane-Shastry model \cite{Haldane1988,Shastry1988} which can be solved by the Bethe ansatz. In addition, we find that a $1/[d^2\sinh^2(x/d)+d^2]$ potential, a non-power-law potential related to the Calogero-Sutherland model \cite{Calogero1969,Sutherland1971}, also can develop a nonmonotonic dispersion as plotted in Fig.~\ref{Fig:other_dispersion}(b). 
The existence of the nonmonotonic dispersion is indeed not particular to the $1/r$ Coulomb interaction and should exist in other systems with long-range interactions (e.g., trapped ions \cite{Blatt2012}). It is possible that the nonmonotonic maxon-roton charge collective mode dispersion is indeed a universal feature of all strongly interacting 1D models with long-range potentials of any type although it is desirable to establish this speculation using exact techniques such as the Bethe ansatz.

\section*{Acknowledgments}
We thank Shiang Fang, An\'ibal Iucci, Rahul Nandkishore, and Micheal Schecter for useful discussions. 
In particular, we thank Rahul Nandkishore for his insightful inputs in the early stage of this work.
This work is supported by the Laboratory for Physical Sciences (Y-.Z.C. and S.D.S.) and by the Army Research Office under Grant Number W911NF-17-1-0482 (Y-Z.C.).
The views and conclusions contained in this document are those of the authors and should not be interpreted as representing the official policies, either expressed or implied, of the Army Research Office or the U.S. Government. The U.S. Government is authorized to reproduce and distribute reprints for Government purposes notwithstanding any copyright
notation herein.




\end{document}